\documentclass{pasj00}
\draft

\begin{document}
\SetRunningHead{Fu-Wen Zhang et al.}{Temporal Properties in
Gamma-ray Bursts}
\Received{}
\Accepted{}

\title{Dependence of Temporal Properties on Energy in
Long-Lag, Wide-Pulse Gamma-Ray Bursts}

\author{Fu-Wen Zhang\altaffilmark{1,4}, Yi-Ping Qin\altaffilmark{2,3}
and Bin-Bin Zhang\altaffilmark{1,4,5}}

\altaffiltext{1}{National Astronomical Observatories/Yunnan
Observatory, Chinese Academy of Sciences, P.O. Box 110, Kunming,
Yunnan 650011, China }

\email{fwzhang@hotmail.com; fwzhang@ynao.ac.cn}

\altaffiltext{2}{Center for Astrophysics, Guangzhou University,
Guangzhou 510006, China}

\altaffiltext{3}{Physics Department, Guangxi University, Nanning,
 Guangxi 530004, China}

\altaffiltext{4}{The Graduate School of the Chinese Academy of
Sciences, P.O. Box 3908, Beijing 100039, China}

\altaffiltext{5}{Department of Physics and Astronomy, University of
Nevada, Las Vegas, NV 89154, USA}


\maketitle

\begin{abstract}
We employed a sample compiled by Norris et al. (2005, ApJ, 625, 324)
to study the dependence of the pulse temporal properties on energy
in long-lag, wide-pulse gamma-ray bursts. Our analysis shows that
the pulse peak time, rise time scale and decay time scale are power
law functions of energy, which is a preliminary report on the
relationships between the three quantities and energy. The power law
indexes associated with the pulse width, rise time scale and decay
time scale are correlated and the correlation between the indexes
associated with the pulse width and the decay time scale is more
obvious. In addition, we have found that the pulse peak lag is
strongly correlated with the CCF lag, but the centroid lag is less
correlated with the peak lag and CCF lag. Based on these results and
some previous investigations, we tend to believe that all
energy-dependent pulse temporal properties may come from the joint
contribution of both the hydrodynamic processes of the outflows and
the curvature effect, where the energy-dependent spectral lag may be
mainly dominated by the dynamic process and the energy-dependent
pulse width may be mainly determined by the curvature effect.

\end{abstract}

\section{Introduction}

Cosmic gamma-ray bursts (GRBs) exhibit a great diversity of the
temporal and spectral structure, and their origin and mechanism are
still unclear. In many bursts the temporal activity is suggestive of
a stochastic process. It was suggested that some simple bursts with
well-separated structure might consist of fundamental units of
emission such as pulses, with some of them being seen to comprise a
fast rise and an exponential decay (FRED) (see, e.g. Fishman 1994).
The temporal and spectral properties of these fundamental pulses
might give us valuable clues about the origin of these events and
will provide powerful constrains on the detailed physical process.

Recently, the temporal and spectral characteristics of GRB pulses
have been intensively studied, and several significant correlations
between them have been found. \citet{nor86} first noted that GRB
pulses have an general observed trend of hard to soft spectral
evolution. This has been confirmed by many other authors (see, e.g.,
Bhat et al. 1994; Norris et al. 1996; Band 1997). The hard-to-soft
spectral evolution are associated with two distinct, observed
features: pulse peaks shift to later times and pulses become wider
at lower energies (e.g., Link, Epstein, \& Priedhorsky 1993; Norris
et al. 1996; Norris, Narani \& Bonnell 2000; Norris et al. 2005,
hereafter Paper I ). By using the average autocorrelation function
and the average pulse width, \citet{fen95} showed that the narrowing
of pulse width with energy well follow a power law, with an index of
$\sim -0.4$. This is the first quantitative relationship between the
temporal and spectral structure in GRBs. \citet{nor96} proposed a
``pulse paradigm" and found that the average raw pulse shape
dependence on energy is also approximately a power law, consistent
with the autocorrelation analysis of \citet{fen95}. This was further
confirmed by later studies \citep
{pir98,cos99,nem00,fer01,cre03,qin04,qin05,pen06}.

In the standard fireball scenario, it was suggested that the process
of radiation in GRBs is most likely through synchrotron emission
(see, e.g., Katz 1994; Sari, Narayan, \& Piran 1996). The power-law
dependence of pulse width on energy has led to the suggestion that
this relationship could be related with synchrotron radiation
\citep{fen95,coh97,pir99}. \citet{kaz98} proposed that the result
could be accounted for by synchrotron cooling (see also Chiang 1998;
Dermer 1998; Wang et al. 2000). It was suspected that the power-law
relationship might result from a relative projected speed or a
relative beaming angle \citep{nem00}. Recently, it has been also
argued that the relativistic curvature effect could lead to the
power-law relationship \citep{qin04,qin05,she05,pen06}.
\citet{dad07} suggested that such correlation is a straightforward
prediction of the `cannonball' model of GRBs \citep{dar04}.

The phenomenon of GRB pulse peaks evolving from higher to lower
energies is a prevalent property of most bursts. Many authors
generally analyze a time delay between the light curves in different
energy bands. Using the cross-correlation method, \citet{che95}
first found that soft emission had a time delay relative to
high-energy emission and quantified the delay. Subsequently, several
investigations on the GRB lag have been carried out
\citep{nor96,nor00,wu00,hak04,hak06,che05,nor06,yi06,zha06b,zha06c}.
There have been several attempts to explain the origin of the time
lag. It was suggested that the activity of the central engine and
hydrodynamic time-scale of the internal shocks might produce the
time lag (e.g. Daigne \& Mochkovitch 1998; 2003; Wu \& Fenimore
2000). \citet{iok01} proposed that the lag was caused by the viewing
angle of the jet. Another possible origin of the lag was proposed to
be the radiative cooling (e.g. Zhang et al. 2002; Bai \& Lee 2003;
Schaefer 2004). \citet{koc03} assumed that the observed lag was the
direct result of spectral evolution (see also Ryde 2005).
\citet{she05} argued that the observed lags could be accounted for
by the curvature effect of fireballs (see also Lu et al. 2006).

\citet{krl03} found that there is a linear relationship between the
pulse rise time and the pulse width (see figure 10 in the paper).
The same result was also found in GRB pulses observed by the
INTEGRAL (see figure 5a in Ryde et al. 2003). Recently, the strong
correlation between the pulse rise time and the pulse width in
different energy channels was presented by authors of Paper I. They
fitted the two quantities with a power-law function and found that
the slope increasing from 0.7 to 1.0 as the energy channel increases
and the correlation is the tightest in channel 3 ($100-300$ keV).
\citet{lqy06} further studied the relationship between the two
quantities and proposed that merely the curvature effect could
reproduce the correlation.

Although a power-law anti-correlation between pulse width and
energy, a strong correlation between pulse rise time and pulse
width, and pulse peaks evolve in time from higher to lower energies
in many GRBs have been studied by many authors, it is unclear how
the pulse peak time, rise time scale and decay time scale depend on
energy. Recently, \citet{lia06} have tentatively investigated the
correlation between the peak time and the average photon energy for
GRB 060218, which has the longest pulse duration and spectral lag
observed to date among the observed GRBs, and found that $t_{peak}
\propto E ^{-0.25\pm0.05}$. It is known that most bright bursts have
many narrow pulses that are difficult to model due to overlapping.
However, the relatively simple, long spectral lag, wide-pulse bursts
are easier to model and might have simpler physics. Since the pulses
in long-lag bursts are very long, the sufficient pulse definition is
available, which makes the study easier. Authors of Paper I have
analyzed the temporal and spectral behavior of wide pulses in 24
long-lag bursts, using a pulse model with two shape parameters,
width and asymmetry, and the Band spectral model with three shape
parameters. They found that the five descriptors are essentially
uncorrelated, but pulse width is strongly correlated with spectral
lag. They also found that pulses in long-lag bursts are
distinguished from those in bright bursts: pulses in long spectral
lag bursts are fewer in number and $\sim100$ times wider (tens of
seconds), have systematically lower peaks in $\nu F_{\nu}$, and have
significantly softer spectra.

As discovered in \citet{nor02}, proportion of long-lag, wide pulses
within long-duration bursts increases from negligible among bright
BATSE bursts to $\sim 50 \%$ at the trigger threshold. Long-lag
bursts appear to be important since these bursts may form a separate
subclass of GRBs \citep{lia07}, and have relatively simple physical
mechanism. Based on the fact that redshifts of three such bursts are
available [GRB 980425 \citep{gal98}, 031203 \citep{mal04} and 060218
\citep{mir06}], it was argued that long-lag bursts are probably
relatively nearby, and the local event rate of these GRBs should be
much higher than that expected from the high luminosity GRBs
\citep{lia07,cob06,pia06,sod06}. It was suggested that their
wide-pulse, long-lag, and under-luminous features are partly
attributed to the off-axis viewing angle effect
\citep{nak99,sal00,iok01}, and partly due to their lower Lorentz
factors \citep{kul98,woo99,sal00,dai06,wan06}. Recently, it was
argued that they might have a different type of central engine (e.g.
neutron stars rather than black holes) from bright GRBs
\citep{maz06,sod06,tom07}.

In this paper, we employ the long-lag burst sample investigated in
Paper I to analyze the dependence of their temporal properties on
energy. We describe the sample and data in section 2. The results
are presented in section 3, followed by conclusions in section 4.

\section{Sample and Data Description}

The GRB sample employed is that presented in Paper I, where the
bursts are found to consist of few long-lag, wide, well-defined
pulses. The data are provided by the BASTE instruments on board the
CGRO spacecraft. In this sample, obvious migration of peaks of the
pulses in different energies can be observed. The bursts of the
sample are from 1429 BATSE events described in \citet{nor02}, with
the criterion that $T_{90}>2$ s, $F_{peak}>0.75$ photons cm$^{-2}$
s$^{-1}$ (50-300 keV), peak intensity PI$>1000$ counts s$^{-1}$
($>25$ keV) and average lag $>1$ s. In addition, only the bursts
with sufficiently non-overlapping pulses are considered. The sample
consists of 24 bursts, most of which contain single pulses. (For
more details of the sample selection, see Paper I.)

For the purpose of fitting a pulse, authors in Paper I developed a
pulse model with a form containing two exponentials, one increasing
and one decreasing with time. This pulse model is written as
$I(t)=A\lambda/[exp(\tau_{1}/t)exp(t/\tau_{2})]$, where
$\lambda=exp[2(\tau_{1}/\tau_{2})^{1/2}]$, $A$ is the pulse peak
intensity, and $\tau_{1}$ and $\tau_{2}$ are the two fundamental
timescales dominating the rise and decay rates, respectively. The
time of pulse onset with respect to $t=0$, $t_{s}$, is ignored. The
24 long-lag bursts were fitted with this model in Paper I. Parameter
values for all identified pulses were obtained, including pulse peak
intensity ($A$), pulse onset time ($t_{s}$), effective onset time
($t_{eff}$), peak time ($\tau_{peak}$), the two fundamental
timescales ($\tau_{1}$ and $\tau_{2}$), width ($w$) and asymmetry
($k$) (see Table 2 in Paper I). The corresponding errors were also
estimated. The effective onset time, $t_{eff}$, is defined as the
time when the pulse reaches 0.01 times of the peak intensity. Both
onset times are relative to the burst trigger time. The peak time is
defined as that relative to the effective onset time.

\section{Results}

Due to a variety of interpretation of the spectral lag observed in
GRBs, we suspect that the quantity might be contributed by various
effects. The most important one might be the mechanism of shocks
which are likely to dominate light curves of pulses in their rise
phase. Another important factor might be the curvature effect which
seems to dominate the decay phase of pulses (see, e.g., Qin \& Lu
2005). We thus pay our attention on how the pulse peak time, rise
and decay time scales depending on energy, checking if the
dependence is the same or different.

\subsection{Dependence of pulse peak time, rise time and decay
time scales on energy}

To investigate this issue, let us define three quantities: pulse
peak time position ($t_{p}$), pulse rise time scale ($\Delta t_{r}$)
and pulse decay time scale ($\Delta t_{d}$), where $t_{p}$ is
defined as the time between the pulse peak and the pulse onset,
$\Delta t_{r}$ and $\Delta t_{d}$ are defined as the time between
the pulse peak and the two $1/e$ intensity points respectively as
those defined in Paper I. (Note that $\Delta t_{r}$ and $\Delta
t_{d}$ are close to the FWHMs in the rising phase and decaying phase
respectively, since $1/e$ is close to 1/2. In addition, according to
their definitions, $\Delta t_{r}=\tau _{rise}$ and $\Delta
t_{d}=\tau_{dec}$, where $\tau_{rise}$ and $\tau_{dec}$ are the
pulse rise and decay timescales defined in Paper I, respectively.)
The onset time, $t'_s$, defined here is the time (relative to the
burst trigger time) when the total counting rate of all four energy
channels ($25-50, 50-100, 100-300$, and $>300$ keV) reaches 0.01
times the peak intensity in a single-pulse burst. With this
definition, the pulse peak time positions, $t_{p}$, in the different
channels for a burst are relative to the same reference time (the
onset $t'_s$). We thus can directly compare them (or, the shifts of
the pulse peaks with respect to the different channels can be easily
estimated). The $\tau_{peak}$ listed in Table 2 of Paper I is
relative to $t_{eff}$ (for a same burst, values of $t_{eff}$ are
different in the different energy channels), which is nothing but a
measure of the pulse rise time as described in Paper I. The peak
time position $t_{p}$ is merely a shift of $\tau_{peak}$ of
individual pulse, but for a burst the shifting steps are different
in the different channels.

We employ all GRBs presented in Paper I to study the dependence of
$t_{p}$, $\Delta t_{r}$ and $\Delta t_{d}$ on energy. For each burst
we require that the pulse signal should be detectable in at least
three channels (in this way, the relationship between these
quantities and energy can be studied). Pulses of the same burst
being blended, such as the three pulses in the burst of \#2711, are
not included. With these requirements, we obtain 24 pulses which
belong to 23 bursts. In the analysis of the relationship between
$t_{p}$ and energy, we only consider the single-pulse bursts since
the onset is well determined. The burst of \#8049 is excluded.

Parameters of the model described by equation (1) in Paper I for the
24 pulses in different energy channels are available in Table 2 of
the paper. According to equation (2) of Paper I, $\Delta t_{r} +
\Delta t_{d}$ ($\Delta t_{r} + \Delta t_{d}= w$) could be determined
by $\tau_1$ and $\tau_2$. We thus obtain $\Delta t_{r}$ and $\Delta
t_{d}$ for the 24 pulses in different energy channels from that
table via a simple derivation. Listed in Table 2 of Paper I is also
the effective onset time $t_{eff}$ defined in that paper, which is
relative to the trigger time, for each pulse in each channel. With
$t_{eff}$ and $\tau_{peak}$ we are able to determine the pulse peak
time relative to the trigger time. We combine the 64 ms count data
from all four channels (the data are available via anonymous ftp in
the website\footnote{
ftp://cossc.gsfc.nasa.gov/compton/data/batse/}) to obtain the
``bolometric" light-curve profile, and derive our onset time $t'_s$
for each burst by fitting the light curve with the method of Paper
I, where the adopted pulse model is equation (1) of the paper.
Shifting $\tau_{peak}$ from $t_{eff}$ to $t'_s$ yields the peak time
$t_p$ defined in this paper. The correspondent uncertainties are
calculated through the error transfer formula.

Illustrated in Figure 1 are $t_{p}$, $\Delta t_{r}$ and $\Delta
t_{d}$ of individual pulses in each energy channel. To compare with
the dependence of the pulse width on energy, the value of $\Delta
t_{w}$, which is defined as the time between the two $1/e$ intensity
points of individual pulse as that defined in Paper I ($\Delta t_{w}
= w$), of each pulse is also displayed in Figure 1. The figure shows
clearly that $t_{p}$ generally migrates to later times at lower
energy channels, and $\Delta t_{w}$, $\Delta t_{r}$ and $\Delta
t_{d}$ become wider at lower energy bands (in several exception
cases, the data points are well inside the corresponding trend
within $1\sigma$ errors).

According to Figure 1, we assume that the four time quantities
$t_{p}$, $\Delta t_{w}$, $\Delta t_{r}$, and $\Delta t_{d}$ are
power law functions of energy. The dependence of the four time
quantities on energy is parameterized by the power law index which
is obtained by fitting the data points of each quantity in the four
energy channels with a power law. The energy adopted for a channel
is the geometric mean of the lower and upper boundaries of the
channel (here we use $300-1000$ keV for channel 4, which is adopted
throughout this paper). This method of analysis was generally
adopted in previous works (see, Fenimore et al. 1995; Norris et al.
1996; Paper I). Let $\alpha_{p}$, $\alpha_{w}$, $\alpha_{r}$, and
$\alpha_{d}$ denote the indices of the power law relationships
between $t_{p}$, $\Delta t_{w}$, $\Delta t_{r}$, and $\Delta t_{d}$
and energy, respectively. Displayed in Figure 2 are the
distributions of these indices. We fit them with a Gaussian. Values
of the fit (the standard deviation) as well as the medians of the
distributions of the four power-law indices are listed in Table 1.
[For the distribution and other analysis of $\alpha_{w}$, see also
Jia \& Qin (2005); Peng et al. (2006).] One can find from Figure 2
and Table 1 that the distributions of these indices have large
dispersions. This implies that the energy dependence of the temporal
properties may not be the same for different bursts. It is
interesting that the distribution of $\alpha_{r}$ is obviously
narrower than that of other indices (see Table 1). A possible
interpretation to this phenomenon is that the mechanism causing the
dependence of the rise time scale on energy might be somewhat
similar for different bursts.

In the analysis of the relationship between the pulse width and
energy, one generally studied the dependence of the average pulse
width on energy for the adopted samples (see, for example, Fenimore
et al. 1995; Norris et al. 1996; Paper I). Here, we also calculate
the dependence of the average values of $t_{p}$, $\Delta t_{r}$ and
$\Delta t_{d}$ on energy. For the sake of comparison, the energy
dependence of the average value of $\Delta t_{w}$ is displayed as
well (we include only those bursts with their pulse signal being
detectable in all four channels; the burst of \#6526, which has a
very long timescale, is excluded throughout the paper). Plotted in
Figure 3 are the relationships between the average values of
$t_{p}$, $\Delta t_{w}$, $\Delta t_{r}$, $\Delta t_{d}$ and energy.
The regression analysis yields: $log
t_{p}=(1.45\pm0.26)-(0.25\pm0.14)log E$, $log \Delta
t_{w}=(2.15\pm0.09)-(0.45\pm0.05)log E$, $log \Delta
t_{r}=(1.40\pm0.10)-(0.37\pm0.05)log E$ and $log \Delta
t_{d}=(2.08\pm0.09)-(0.48\pm0.05)log E$.

\subsection{Relationships between power-law indices}

As power law indices are an active factor reflecting the
relationship between the temporal and spectral properties of pulses,
we are curious about how the three power-law indices, $\alpha_{w}$,
$\alpha_{r}$ and $\alpha_{d}$, which are associated with various
widthes of pulses, are related. Figure 4 shows the relations between
them. Results of the correlation analysis for the three quantities
are listed in Table 2. We find that $\alpha_{w}$ and $\alpha_{d}$
are highly correlated, while the other pairs of the quantities are
obvious less correlated. It suggests that the mechanism causing the
power law relationship between the pulse width and energy is the
same as that between the pulse decay time scale and energy. Recall
that, the distributions of these indices have large dispersions
which implies that the energy dependence of these temporal
properties may not be the same for different bursts. We guess that
the energy dependence of the rise time scale and that of the decay
time scale for the same burst during the same pulse might share some
mechanism which is unclear currently. If this mechanism varies from
burst to bursts, there would exist a weak correlation between
$\alpha_{r}$ and $\alpha_{d}$ as observed in Figure 4.

Shown in other aspects, correlation analysis between $\Delta t_{r}$,
$\Delta t_{d}$ and $\Delta t_{w}$ in different energy channels might
be helpful. The results are illustrated in Figure 5 which shows that
$\Delta t_{r}$, $\Delta t_{d}$ and $\Delta t_{w}$ are correlated and
the strong correlations between $\Delta t_{d}$ and $\Delta t_{w}$
exist in each of the three energy channels. This is consistent with
the previous studies \citep{krl03,ryd03,lqy06}. What hinted and
concluded by the correlation analysis of the indices are reinforced
by these new results. One should keep in mind that correlations
between different temporal properties might partially (or mainly) be
due to the same Lorentz factor for the same pulse (see, Lu, Qin \&
Yi 2006), but the more obvious correlation between $\Delta t_{d}$
and $\Delta t_{w}$ than that between other pairs suggests that,
besides the Lorentz factor, there must be other factors at work in
producing the strong correlation between the two quantities.  As
shown in Zhang \& Qin (2005), the ratio of $FWHM_r$ to $FWHM_d$ is
not affected by the Lorentz factor.

One might notice that, in terms of mathematics, the strong
correlations between $\alpha_{d}$ and $\alpha_{w}$ and between
$\Delta t_{d}$ and $\Delta t_{w}$ may result from the fact that
$\Delta t_{w}$ is dominated by $\Delta t_{d}$. Or, in turn, the
strong correlations between $\alpha_{d}$ and $\alpha_{w}$ and
between $\Delta t_{d}$ and $\Delta t_{w}$ may confirm the fact. As
shown in \citet{qin04}, the ratio of $FWHM_r$ to $FWHM_d$ would be
less than $1.3$ for pulses arising from the emission of
relativistically expanding fireballs. Therefore, it is expected that
$\Delta t_{w}$ might generally be dominated by $\Delta t_{d}$, as
what suggested in Figure 1.

From Figures 2 and 4 one finds that $\alpha_{d}< \alpha_{r}$. This
suggests that the decay time scale rapidly decreases with respect to
energy, while the variance of the rise time scale with the
increasing of energy is relatively mild. Is it implying that the
curvature effect plays an important role in the decaying phase of
pulses and the contribution of the effect makes $\alpha_{d}$ smaller
(see, e.g., Qin et al. 2005; Peng et al. 2006)?

\subsection{Relationships between various spectral lags and between the lags and other time scales}

Authors of Paper I measured the peak lags of all pulses between
channels 2 and 3 in 24 long-lag bursts, and found that as pulse
width increases, the spectral lag measured between pulse peaks tends
to increase. We find that not only the peak time lag (note that what
we measure here is the peak lag between channels 1 and 3,
$\tau_{p,13}$) but also the CCF lag, which is the lag calculated
with the cross correlation function (CCF) method, increase with the
increasing of the pulse width (the figure is omitted). The CCF lag
used here is also derived between channels 1 and 3, $\tau_{CCF,13}$,
which has been extensively studied \citep
{lin93,che95,nor96,nor00,wu00,hak04,hak06,che05,nor06,yi06,zha06b,zha06c}.
Here, we derive the CCF lag from the peak of the CCF without
considering the side lobe contribution of the CCF. Since the light
curves are the smooth pulses and their lags are significantly larger
than the time bin, the peaks of CCFs are robust to estimate the
lags. The errors of CCF lags are evaluated by simulations. Besides
these two lags, the centroid lag which is the lag of the pulse
centroid was discussed in Paper I, and it was found to be well
measured and to be well correlated with the pulse width. It was
suggested recently that the correlation might be due to the Lorentz
factor (see, e.g. Peng et al. 2007).

To analyze the relationships between the three lags, we calculate
the centroid lag between channels 1 and 3 ($\tau_{cen,13}$) as well.
The plots of $\tau_{cen,13}$ vs. $\tau_{p,13}$, $\tau_{cen,13}$ vs.
$\tau_{CCF,13}$, and $\tau_{CCF,13}$ vs. $\tau_{p,13}$ are displayed
in Figure 6. One finds that  $\tau_{cen,13}$ is weakly correlated
with both $\tau_{p,13}$ and $\tau_{CCF,13}$, while the later two are
strongly correlated. The best fits to $\tau_{CCF,13}$ and
$\tau_{p,13}$ yields
$log\tau_{CCF,13}=(-0.25\pm0.06)+(1.18\pm0.11)log \tau_{p,13}$. The
strong correlation between $\tau_{p,13}$ and $\tau_{CCF,13}$ and the
weak correlations between the two quantities and $\tau_{cen,13}$
suggest that $\tau_{CCF,13}$ is mainly caused by the shifting of
peaks while $\tau_{cen,13}$ is not. We believe that $\tau_{p,13}$
and $\tau_{cen,13}$ reflect different aspects of spectral lags, with
one representing the shifting of peaks and the other describing the
enhancement of the time scale of pulses. We thus propose that, to
reveal a spectral lag in detail, both $\tau_{p,13}$ and
$\tau_{cen,13}$ should be measured.

In addition, we find that $\tau_{cen,13}$ is systematically larger
than both $\tau_{p,13}$ and $\tau_{CCF,13}$. According to the above
interpretation, this implies that the lag caused by the stretching
of pulses is always larger than that caused by the shifting of
peaks.

\citet{hak06} found that GRB lags are consistent across a wide range
of prompt emission energies, $lag_{31}\approx lag_{21}+lag_{32}$.
Under the interpretation proposed above, the three lags, $lag_{31}$,
$lag_{21}$ and $lag_{32}$ are mainly due to the shifting of $t_p$ in
the corresponding channels. Therefore, they could be approximated by
$\tau_{p,13}$, $\tau_{p,12}$ and $\tau_{p,23}$, respectively.
Meanwhile, according to their definitions, one has
$\tau_{p,13}=\tau_{p,12}+\tau_{p,23}$. The relation $lag_{31}\approx
lag_{21}+lag_{32}$ is thus explained.

\citet{bha94} found that the time lag between the counting rate and
the hardness ratio was directly correlated with the rise time of the
burst counting rate profile. Motivated by this, we analyze the
relationships between the three lags and the pulse rise time and
decay time scales. The results are displayed in Figure 7. It shows
that the peak lag and CCF lag are well correlated with the pulse
rise time scale and weakly correlated with the pulse decay time
scale, which is consistent with that found by \citet{pen07}.
However, the centroid lag is strongly correlated with the pulse
decay time scale and weakly correlated with the pulse rise time
scale. The latter phenomenon is in agreement with what interpreted
above. As discussed in last subsection, $\Delta t_{w}$ is likely
dominated by $\Delta t_{d}$. Thus, it is expectable that
$\tau_{cen,13}$ is correlated with $\Delta t_{d}$, since according
to the interpretation, the centroid lag reflects the stretching of
the pulse width. The correlations between the peak and CCF lags and
the pulse rise time scale indicate that the two lags might be caused
by some mechanism associated with the pulse rise time scale.
Probably, the peak and CCF lags and the pulse rise time scale might
be created mainly by a dynamic process, while the centroid lag and
the pulse decay time scale might be formed by both the dynamic
process and the curvature effect.

\section{Conclusions} \label{con}

Using the sample of 24 long-lag, wide-pulse GRBs described in Paper
I, we have investigated the dependence of the pulse temporal
properties on energy. It is obvious that the peak time generally
migrates to later time at lower energy channels, and the pulse
width, rise time and decay time scales become wider at lower energy
bands. Fitting the average pulse peak time, rise time and decay time
scales with a power law function of energy yields $t_{p}\propto
E^{-0.25\pm0.14}$, $t_{r}\propto E^{-0.37\pm0.05}$ and $t_{d}\propto
E^{-0.48\pm0.05}$. This is a preliminary report on the relationships
between the three quantities and energy. The three power law indices
$\alpha_{p}$, $\alpha_{r}$ and $\alpha_{d}$ have large dispersions,
and the medians of their distributions are $-0.27$, $-0.35$ and
$-0.37$, respectively. It is not surprising since in the well
defined power law relationship between the pulse width and energy
one also finds a large dispersion of the index (see also Jia \& Qin
2005; Peng et al. 2006). This implies that the energy dependence of
the temporal properties may not be the same for different bursts. It
is interesting that the distribution of $\alpha_{r}$ is obviously
narrower than that of other indices (see Table 1). A possible
interpretation to this phenomenon is that the mechanism causing the
dependence of the rise time scale on energy might be somewhat
similar for different bursts. \citet{lia06} noted that the peak time
dependence on the average energy (from 0.3-150 keV) in the single
pulse burst GRB 060218 detected by Swift was approximately a power
law, and the power law index was $\sim -0.25\pm0.05$, which is
consistent with our result. This favors what argued by \citet{lia06}
that this event may be a typical long-lag, wide-pulse burst and
share the similar radiation physics with other BATSE bursts.

We also find that the three power-law indices $\alpha_{w}$,
$\alpha_{r}$ and $\alpha_{d}$ are correlated, where $\alpha_{w}$ and
$\alpha_{d}$ are found to be more obviously correlated. It suggests
that the mechanism causing the power law relationship between the
pulse width and energy is the same as that between the pulse decay
time scale and energy. Recalling that the distributions of these
indices have large dispersions, implying that the energy dependence
of these temporal properties may not be the same for different
bursts, we guess that the energy dependence of the rise time scale
and that of the decay time scale for the same burst during the same
pulse might share some mechanism which is unclear currently. If this
mechanism varies from burst to bursts, there would exist a weak
correlation between $\alpha_{r}$ and $\alpha_{d}$ as observed in
Figure 4.

In addition, we find that the pulse peak lag is strongly correlated
with the CCF lag, but the centroid lag is weakly correlated with the
peak lag and CCF lag. This suggests that the CCF lag is mainly
caused by the shifting of peaks while the centroid lag is not. We
argue that the peak lag and the centroid lag reflect different
aspects of spectral lags, with one representing the shifting of
peaks and the other describing the enhancement of the time scale of
pulses. We thus propose that, to reveal a spectral lag in detail,
both the peak lag and the centroid lag should be measured. Our
analysis also shows that the centroid lag is systematically larger
than both the peak and CCF lags. According to the above
interpretation, this implies that the lag caused by the stretching
of pulses is always larger than that caused by the shifting of
peaks. According to the definition of the pulse peak lag and the
relation between the peak time and energy, one has
$\tau_{p,13}=\tau_{p,12}+\tau_{p,23}$. Along with the relationship
between the peak lag and CCF lag, the relation $lag_{31}\approx
lag_{21}+lag_{32}$ found by \citet{hak06} can be explained.

According to \citet{ryd02}, the simplest scenario accounting for the
observed GRB pulses is to assume an impulsive heating of the leptons
and a subsequent cooling and emission. In this scenario, the rising
phase of the pulse, which is referred to as the dynamic time (the
crossing time), arises from the energizing of the shell, while the
decay phase is due to geometric and relativistic effects in an
outflow with a Lorentz factor of $\Gamma\gtrsim100$. An intuitive
speculation is that the dependence of the pulse rise time on energy
is attributed to hydrodynamic processes. In the internal shock model
of GRB pulses, there are three contributors to the pulse temporal
structure: cooling, hydrodynamics, and angular spreading timescales
\citep{pir99,pir05,mes02,mes06,zha04}. Thus, the resulting time
profile is a convolution of the three processes. Based on the
current model which requires a much stronger magnetic field and thus
leads to very fast cooling, the typical cooling timescale ($\sim
10^{-6}$ s, see Wu \& Fenimore 2000) is much shorter than the
observed pulse delays, and hence the cooling timescale can not
dominates the pulse profile. The effect of the angular time arising
from kinematics, the so-called curvature effect, on the
characteristics of pulses has been intensively studied
\citep{pan02,qin02,ryd02,krl03,der04,dyk05,zha06a}. It was argued
that the curvature effect might be responsible for the spectral lag
\citep{sal00,iok01,she05,ryd05,lu06}. The relationship between the
pulse width and energy could also be accounted for by the curvature
effect \citep{qin04,qin05,pen06}. However, \citet{she05} found that
the curvature causes an energy-dependent pulse width distribution
but the energy dependence of the width they obtained was much weaker
than the observed $W\propto E^{-0.4}$ one. \citet{yi06} also argued
that the curvature effect alone could not explain the difference of
the spectral lags (see also Shen, Song, \& Li 2005; Lu et al. 2006).
Daigne \& Mochkovitch(1998, 2003) developed a model in the framework
of internal shock model \citep{ree94} and found that if GRB pulses
were produced by internal shocks, their temporal and spectral
properties were probably governed by the hydrodynamics of the flow
rather by the geometry of the emitting shells. Recently,
\citet{lu07} tentatively analyzed the origination of GRB pulses and
found that the decay phase of the observed pulse originates from the
contributions of both the curvature effect and the width of the
intrinsic pulse, and the rising phase of the observed pulses only
comes from the width of the intrinsic pulse (here the width of the
intrinsic pulse is referred to as the dynamic time). We argue that
all energy-dependent pulse temporal properties discussed above might
probably come from the joint contribution of both the hydrodynamic
processes of the outflows and the curvature effect, where the
energy-dependent spectral lag may be mainly dominated by the dynamic
process and the energy-dependent pulse width may be mainly
determined by the curvature effect.

We appreciate the anonymous referee for her/his helpful suggestions.
We thank Jinming Bai and Enwei Liang for their helpful discussions.
This work is supported by National Natural Science Foundation of
China (No. 10573030 and No. 10463001).

\clearpage

\begin{figure}
  \begin{center}
    \FigureFile(160mm,150mm){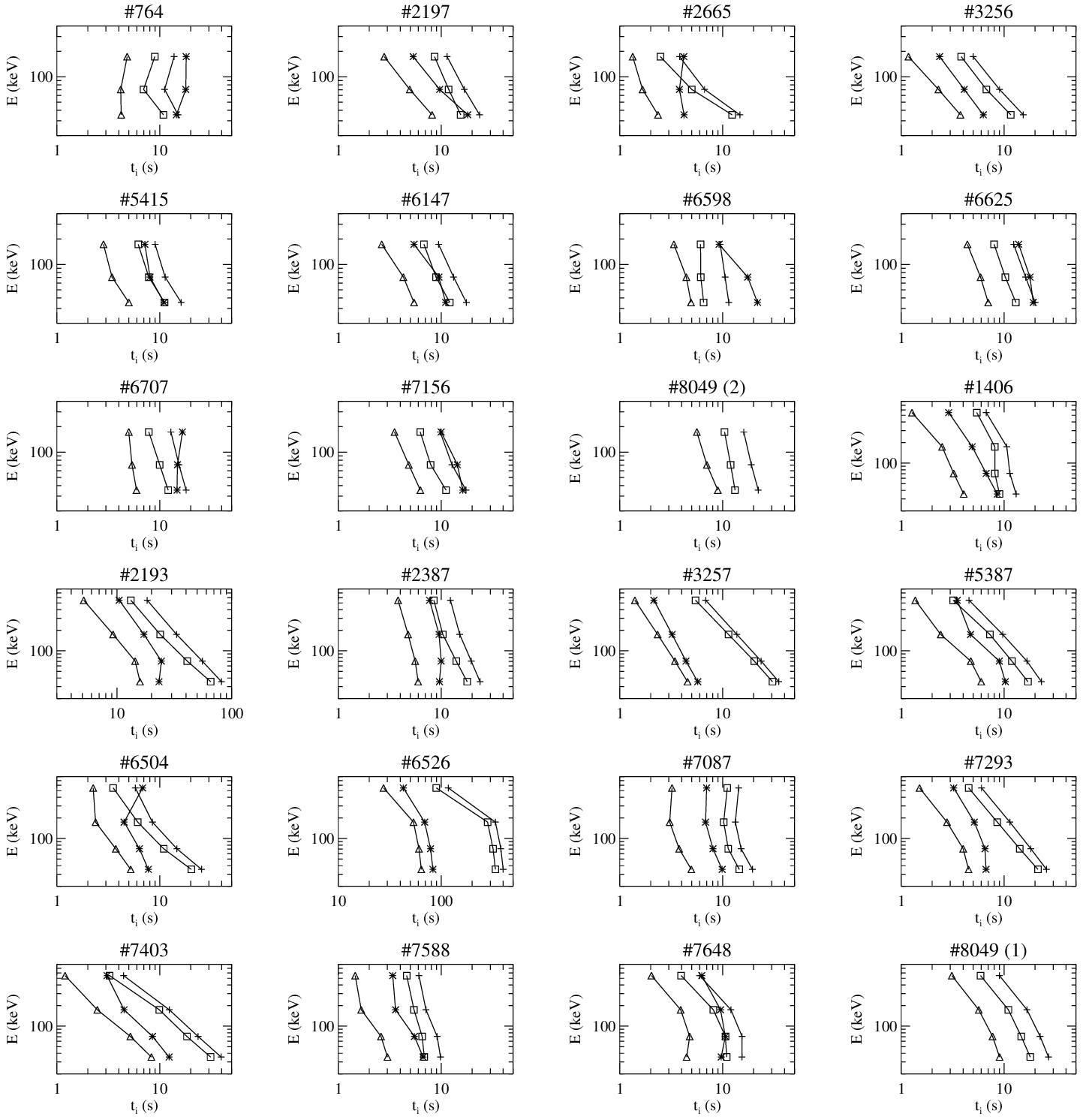}
  \end{center}
  \caption{Energy vs. pulse peak time
(asterisk), pulse width (cross), pulse rise time scale (open
triangle) and pulse decay time scale (open square) for all the 24
pulses studied in this paper, where $E$ is the geometric means of
the lower and upper channel boundaries, $t_{i}$ represents $t_{p}$,
$\Delta t_{w}$, $\Delta t_{r}$ and $\Delta t_{d}$. Symbols joined by
line segments correspond to the same time quantity in the different
energy channels.}\label{fig1}
\end{figure}

\begin{figure}
  \begin{center}
    \FigureFile(120mm,120mm){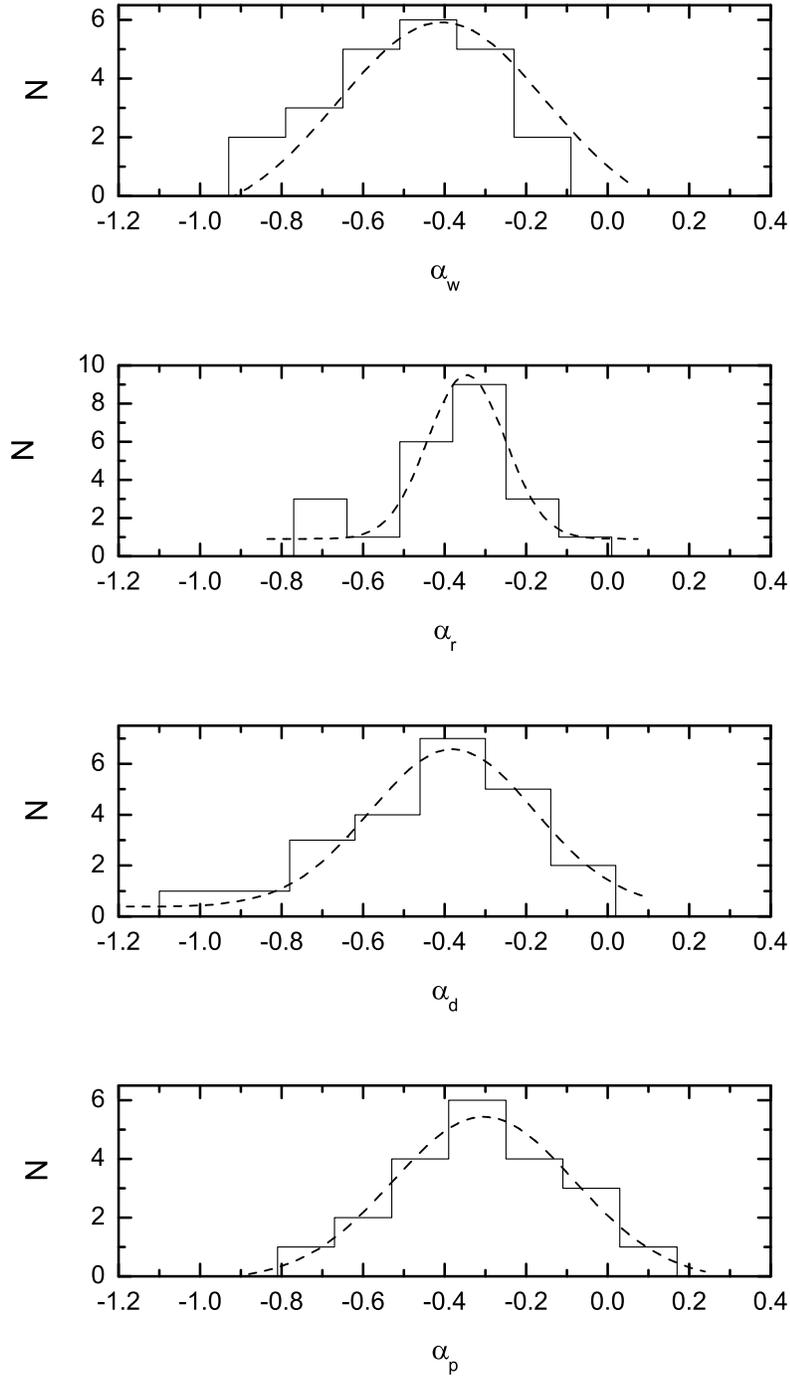}
  \end{center}
  \caption{Distributions of the
power-law indices $\alpha_{w}$ (the first panel), $\alpha_{r}$ (the
second panel), $\alpha_{d}$ (the third panel) and $\alpha_{p}$ (the
fourth panel) obtained by fitting the pulse width, rise time scale,
decay time scale, and peak time and energy with power law functions,
respectively. The dashed lines are the best fits by the Gaussian
functions.}\label{fig2}
\end{figure}

\begin{figure}
  \begin{center}
    \FigureFile(80mm,80mm){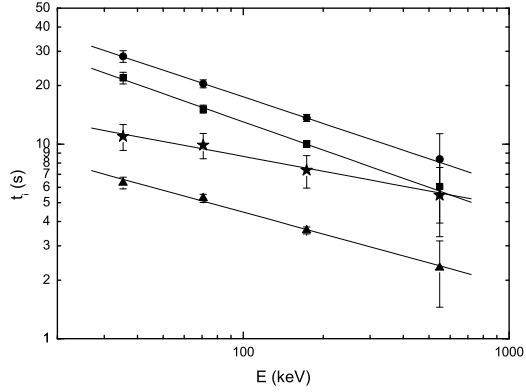}
  \end{center}
  \caption{The average pulse peak time
(filled pentagon), width (filled circle), rise time scale (filled
triangle) and decay time scale (filled square) as the functions of
energy, where $t_{i}$ represents the average values of $t_{p}$,
$\Delta t_{w}$, $\Delta t_{r}$ and $\Delta t_{d}$. The solid lines
are the best fits.}\label{fig3}
\end{figure}

\begin{figure}
  \begin{center}
    \FigureFile(90mm,100mm){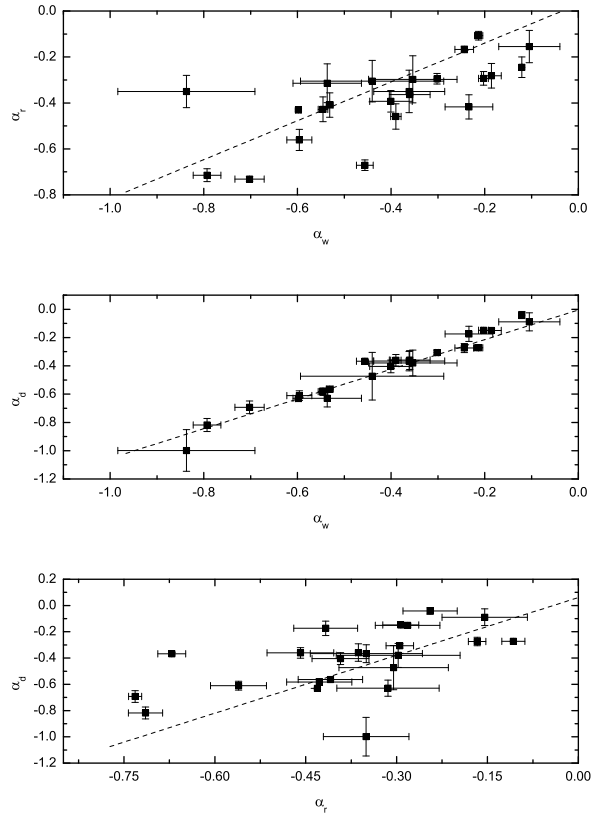}
  \end{center}
  \caption{Relationships between the
three power-law indices $\alpha_{r}$, $\alpha_{d}$ and $\alpha_{w}$.
The dashed lines are the regression lines.}\label{fig4}
\end{figure}

\begin{figure}
  \begin{center}
    \FigureFile(140mm,140mm){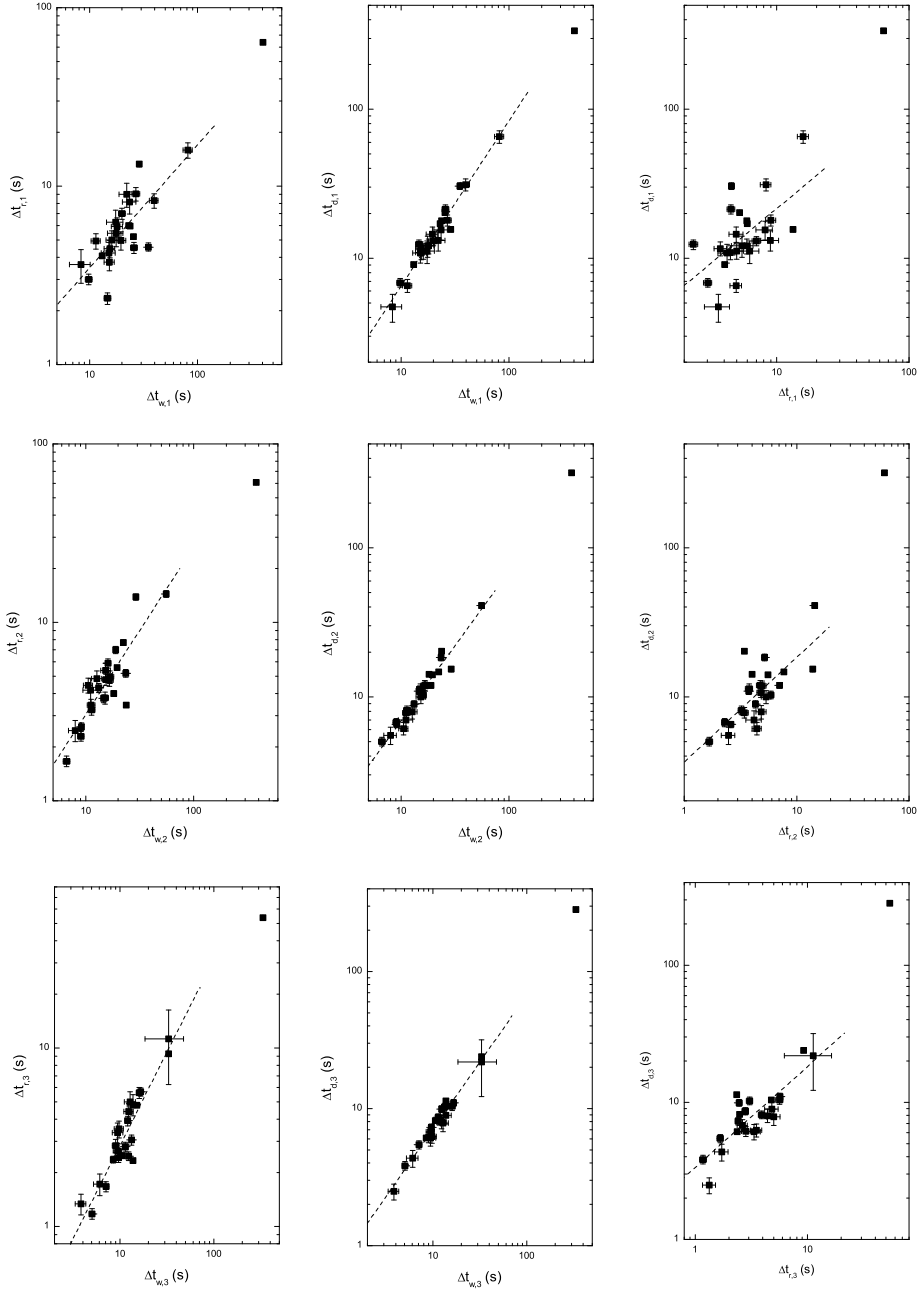}
  \end{center}
  \caption{Plots of $\Delta t_{r}$ vs.
$\Delta t_{w}$, $\Delta t_{d}$ vs. $\Delta t_{w}$ and $\Delta t_{d}$
vs. $\Delta t_{r}$ in the first three energy channels, where
subscript 1, 2 and 3 represent the first channel (the first row),
the second channel (the second row) and the third channel (the third
row), respectively. The dashed lines are the best fits. The $\Delta
t_{r}$ are well correlated with the $\Delta t_{w}$ in channels 1, 2,
and 3 with the slopes of 0.69, 0.94 and 1.05, and R =0.75, 0.87,
0.92, respectively. The $\Delta t_{d}$ and $\Delta t_{w}$ are
strongly correlated with the slopes of 1.11, 1.01 and 0.98 and R
=0.98, 0.98, 0.99 for channels 1, 2, and 3, respectively. There
exist the relatively weak correlations between $\Delta t_{d}$ and
$\Delta t_{r}$ in channels 1, 2, and 3 with the slopes of 0.74, 0.70
and 0.74, and R =0.60, 0.74, 0.84, respectively.}\label{fig5}
\end{figure}

\begin{figure}
  \begin{center}
    \FigureFile(120mm,120mm){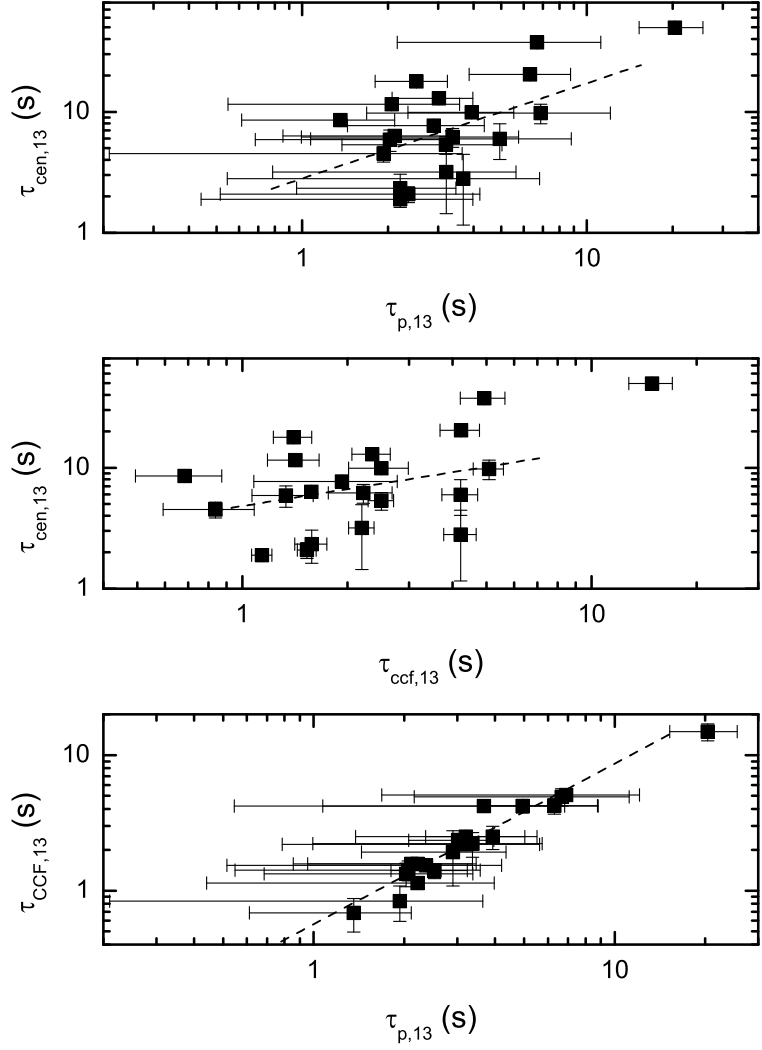}
  \end{center}
  \caption{Plots of the centroid lag
($\tau_{cen,13})$ vs. the peak lag ($\tau_{p,13}$), the centroid lag
vs. CCF lag ($\tau_{CCF,13}$) and the CCF lag vs. the peak lag. The
dashed lines are the best fits, where the correlation coefficients
from the top to bottom panels are 0.44, 0.34, 0.93,
respectively.}\label{fig6}
\end{figure}

\begin{figure}
  \begin{center}
    \FigureFile(140mm,140mm){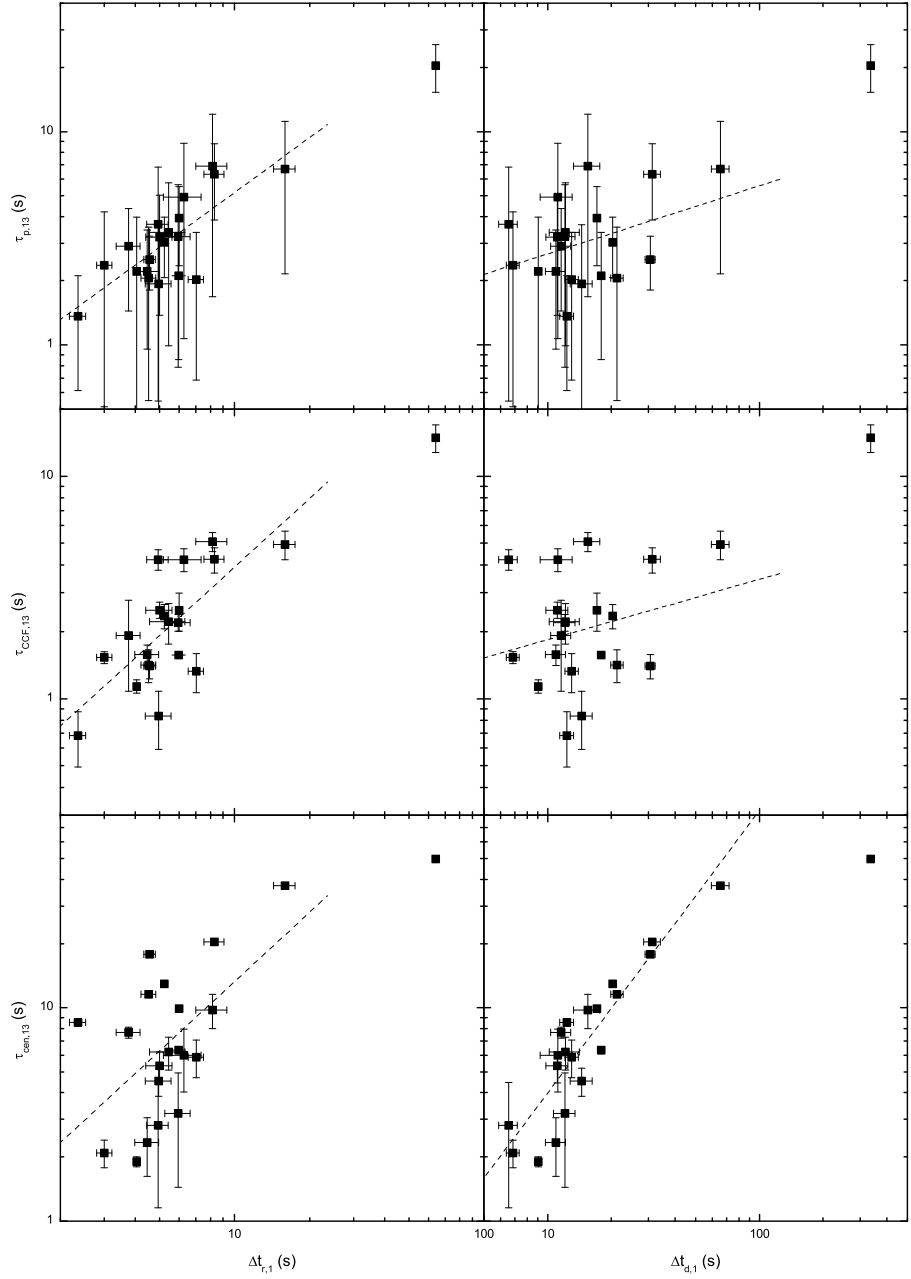}
  \end{center}
  \caption{Relationships between the
three lags and the pulse rise time and decay time scales, where
$\Delta t_{r, 1}$ and $\Delta t_{d, 1}$ are the pulse rise time and
decay time scales in channels 1, respectively. The dashed lines are
the best fits, where the correlation coefficients of between the
peak lag , CCF lag and centroid lag and the pulse rise time scale
are 0.76, 0.71, 0.54 (the first column), and the pulse decay time
scale are 0.39, 0.25, 0.90 (the second column),
respectively.}\label{fig7}
\end{figure}

\clearpage
\begin{table}

\caption{Characteristics of the distributions of the four power-law
indices.\label{tbl-1}}
\begin{center}
\begin{tabular}{ccc}
\hline \hline
 Power-law index & Median  &
 $\sigma$ (modeled with a Gaussian)\\
\hline
 $\alpha_{p}$&$-0.27\pm0.04$ & $0.45\pm0.08$\\
 $\alpha_{w}$&$-0.39\pm0.04$ & $0.51\pm0.11$\\
 $\alpha_{r}$&$-0.35\pm0.03$ & $0.19\pm0.02$\\
 $\alpha_{d}$&$-0.37\pm0.06$ & $0.40\pm0.06$\\
\hline
\end{tabular}

\end{center}
\end{table}

\begin{table}
\begin{center}
\caption{Correlations of the three power-law indices.\label{tbl-3}}
\begin{tabular}{ccc}
\hline\hline
 Correlation& Spearman correlation coefficient & Probability \\
& (r) & (p) \\
\hline
$\alpha_{r}$=(0.03$\pm$0.02)+(0.85$\pm$0.03)$\alpha_{w}$ &0.77 &$5.1\times10^{-5}$\\
$\alpha_{d}$=(-0.01$\pm$0.01)+(1.05$\pm$0.03)$\alpha_{w}$&0.98 &$2.2\times10^{-14}$\\
$\alpha_{d}$=(0.06$\pm$0.03)+(1.47$\pm$0.07)$\alpha_{r}$ &0.66 &$1.1\times10^{-3}$\\
\hline
\end{tabular}

\end{center}
\end{table}


\begin{thebibliography}{}
\bibitem[Bai \& Lee(2003)]{bai03}Bai, J. M., \& Lee, M. G. 2003, \apj, 585, L113
\bibitem[Band(1997)]{ban97} Band, D. L. 1997, \apj, 486, 928
\bibitem[Bhat et al.(1994)]{bha94} Bhat, P. N., et al. 1994, \apj, 426, 604
\bibitem[Chen et al.(2005)]{che05} Chen, L., Lou, Y. Q., Wu, M., Qu,
     J.-L., Jia, S.-M., \& Yang, X.-J. 2005, \apj, 619, 983
\bibitem[Cheng et al.(1995)]{che95} Cheng, L. X., Ma, Y. Q., Cheng,
     K. S., Lu, T., \& Zhou Y. Y. 1995, \aap, 300, 746
\bibitem[Chiang(1998)]{chi98} Chiang, J., 1998, \apj, 508, 752
\bibitem[Cobb et al.(2006)]{cob06} Cobb, B. E., Bailyn, C. D., van Dokkum,
     P. G., \& Natarajan, P. 2006, \apj, 645, L113
\bibitem[Cohen et al.(1997)]{coh97} Cohen, E., Katz, J. I., Piran,
     T., \& Sari, R. 1997, \apj, 488, 330
\bibitem[Costa(1999)]{cos99} Costa, E. 1999, Nucl. Pphs. B (Proc.
     Suppl.), 69, 646
\bibitem[Crew et al.(2003)]{cre03} Crew, G. B., et al. 2003, \apj,
     599, 387
\bibitem[Dado, Dar \& De R\'{u}jula(2007)]{dad07} Dado, S., Dar,
     A., \& De R\'{u}jula, A. 2007, \apj, 663, 400
\bibitem[Dar \& De R\'{u}jula(2004)]{dar04} Dar, A., \& De
     R\'{u}jula, A. 2004, Physics Reports, 405, 203
\bibitem[Dai, Zhang \& Liang(2006)]{dai06}Dai, Z. G, Zhang, B., \& Liang,
     E. W. 2006, preprint (astro-ph/0604510)
\bibitem[Daigne \& Mochkovitch(1998)]{dai98}Daigne, F., \& Mochkovitch,
     R. 1998, \mnras, 296, 275
\bibitem[Daigne \& Mochkovitch(2003)]{dai03}Daigne, F., \& Mochkovitch,
     R. 2003, \mnras, 342, 587
\bibitem[Dermer(1998)]{der98} Dermer, C. D. 1998, \apj, 501, L157
\bibitem[Dermer(2004)]{der04} Dermer, C. D., 2004, \apj, 614, 284
\bibitem[Dyks, Zhang \& Fan(2005)]{dyk05} Dyks, J., Zhang, B., \& Fan, Y.
    Z. 2005, submitted (astro-ph/0511699)
\bibitem[Fenimore et al.(1995)]{fen95} Fenimore, E. E., in 't Zand, J. J. M.,
    Norris, J. P., Bonnell, J. T., \& Nemiroff, R. J. 1995, \apj, 448, L101
\bibitem[Feroci et al.(2001)]{fer01} Feroci, M., et al. 2001, \aap,
     378, 441
\bibitem[Fishman(1994)]{fis94} Fishman, G. J., et al. 1994, \apjs,
     92, 229
\bibitem[Galama et al.(1998)]{gal98} Galama, T. J., et al. 1998, Nature, 395, 670
\bibitem[Hakkila \& Giblin(2004)]{hak04} Hakkila, J., \& Giblin, T.
     W. 2004, \apj, 610, 361
\bibitem[Hakkila \& Giblin(2006)]{hak06} Hakkila, J., \& Giblin, T.
     W. 2006, \apj, 646, 1086
\bibitem[Ioka \& Nakamura(2001)]{iok01} Ioka, K., \& Nakamura, T. 2001,
     \apj, 554, L163
\bibitem[Jia \& Qin(2005)]{jia05} Jia, L.-W., \& Qin, Y.-P. 2005,
     \apj, 631, L25
\bibitem[Katz(1994)]{kat94} Katz, J. I. 1994, \apj, 422, 248
\bibitem[Kazanas, Titarchuk, \& Hua(1998)]{kaz98} Kazanas, D.,
     Titarchuk, L. G., \& Hua, X. M. 1998, \apj, 493, 708
\bibitem[Kocevski \& Liang(2003)]{koc03} Kocevski, D., \& Liang, E. 2003,
     \apj, 594, 385
\bibitem[Kocevski, Ryde \& Liang(2003)]{krl03} Kocevski, D., Ryde, F.,
     \& Liang, E. 2003, \apj, 596, 389
\bibitem[Kulkarni et al.(1998)]{kul98} Kulkarni, S. R., et al. 1998, Nature,
     395, 663
\bibitem[Liang et al.(2006)]{lia06} Liang E. W., Zhang, B. B., Stamatikos, M.,
     Zhang, B., Norris, J., Gehrels, N., Zhang, J., \& Dai, Z. G. 2006,
     \apj, 653, L81
\bibitem[Liang et al.(2007)]{lia07} Liang, E. W., Zhang, B., Virgili, F.,
     \& Dai, Z. G. 2007, \apj, 662, 1111
\bibitem[Link, Epstein, \& Priedhorsky(1993)]{lin93} Link, B., Epstein,
     R. I., \& Priedhorsky, W. C. 1993, \apj, 408, L81
\bibitem[Lu, Qin \& Yi(2006)]{lqy06} Lu, R.-J., Qin, Y.-P., \& Yi,
     T.-F. 2006, ChJAA, 6, 52
\bibitem[Lu et al.(2007)]{lu07} Lu, R.-J., Qin, Y.-P., \& Zhang, F.-W. 2007, ChPhy, 16, 1806
\bibitem[Lu et al.(2006)]{lu06} Lu, R.-J., Qin, Y.-P., Zhang, Z.-B.,
     \& Yi, T.-F. 2006, \mnras, 367, 275
\bibitem[Malesani et al.(2004)]{mal04}Malesani, D., et al. 2004, \apj, 609, L5
\bibitem[Mazzali et al.(2006)]{maz06} Mazzali, P. A. et al. 2006, Nature,
     442, 1018
\bibitem[M\'{e}sz\'{a}ros(2002)]{mes02} M\'{e}sz\'{a}ros, P.
    2002, ARA\&A, 40, 137
\bibitem[M\'{e}sz\'{a}ros(2006)]{mes06} M\'{e}sz\'{a}ros, P.
    2006, Rep. Prog. Phys., 69, 2259
\bibitem[Mirabal et al.(2006)]{mir06} Mirabal, N., Halpern, J. P., An, D., Thorstensen,
    J. R., \& Terndrup, D. M. 2006, \apj, 643, L99
\bibitem[Nakamura(1999)]{nak99} Nakamura, T. 1999, \apj, 522, L101
\bibitem[Nemiroff(2000)]{nem00} Nemiroff, R. J. 2000, \apj, 544, 805
\bibitem[Norris et al.(1986)]{nor86} Norris, J. P., Share, G. H., Messina, D. C.,
   Dennis, B. R., Desai, U. D., Cline, T. L., Matz, S. M., \& Chupp, E. L. 1986, \apj,
     301, 213
\bibitem[Norris et al.(1996)]{nor96} Norris, J. P., Nemiroff, R. J., Bonnell, J. T.,
   Scargle, J. D., Kouveliotou, C., Paciesas, W. S., Meegan, C. A., \& Fishman, G. J.
   1996, \apj, 459, 393
\bibitem[Norris, Narani \& Bonnell(2000)]{nor00} Norris, J. P.,
     Marani, G. F., \& Bonnell, J. T. 2000, \apj, 534, 248
\bibitem[Norris(2002)]{nor02} Norris, J. P. 2002, \apj, 579, 386
\bibitem[Norris et al.(2005)]{nor05} Norris, J. P., Bonnell, J. T., Kazanas, D., Scargle, J. D,
     Hakkila, J., \& Giblin, T. W. 2005, \apj, 627, 324 (Paper I)
\bibitem[Norris \& Bonnell(2006)]{nor06} Norris, J. P., \& Bonnell,
     J. T. 2006, \apj, 643, 266
\bibitem[Panaitescu \& Kumar(2002)]{pan02} Panaitescu, A., \& Kumar, P. 2002, \apj, 571, 779
\bibitem[Peng et al.(2006)]{pen06} Peng, Z.-Y., Qin, Y.-P., Zhang, B.-B., Lu, R.-J.,
    Jia, L.-W., \& Zhang, Z.-B. 2006, \mnras, 368, 1351
\bibitem[Peng et al.(2007)]{pen07} Peng, Z.-Y., Lu, R.-J., Qin, Y.-P., \& Zhang B.-B.
    2007, ChJAA, 7, 428
\bibitem[Pian et al.(2006)]{pia06} Pian, E., et al. 2006, Nature, 442, 1011
\bibitem[Piran(1999)]{pir99} Piran, T. 1999, Phys. Rep., 314, 575
\bibitem[Piran(2005)]{pir05} Piran, T. 2005, Rev. Mod. Phys., 76, 1143
\bibitem[Piro et al.(1998)]{pir98} Piro, L., et al. 1998, \aap, 329,
     906
\bibitem[Qin(2002)]{qin02} Qin, Y.-P. 2002, \aap, 396, 705
\bibitem[Qin et al.(2004)]{qin04} Qin, Y.-P., Zhang, Z.-B., Zhang,
     F.-W., \& Cui, X.-H. 2004, \apj, 617, 439
\bibitem[Qin et al.(2005)]{qin05} Qin, Y.-P., Dong, Y.-M., Lu, R.-J., Zhang, B.-B.,
    \& Jia, L.-W. 2005, \apj, 632, 1008
\bibitem[Qin \& Lu(2005)]{qinl05} Qin, Y.-P., \& Lu, R.-J. 2005, \mnras, 362, 1085
\bibitem[Rees \& M\'{e}sz\'{a}ros(1994)]{ree94} Rees, M. J., \& M\'{e}sz\'{a}ros,
    P. 1994, ApJ, 430, L93
\bibitem[Ryde \& Petrosian(2002)]{ryd02} Ryde, F., \& Petrosian, V.
    2002, \apj, 578, 290
\bibitem[Ryde et al.(2003)]{ryd03}Ryde F., Borgonovo, L., Larsson, S., Lund, N.,
    von Kienlin, A., \& Lichti, G. 2003, \aap, 411, L331
\bibitem[Ryde(2005)]{ryd05} Ryde, F. 2005, \aap, 429, 869
\bibitem[Salmonson(2000)]{sal00} Salmonson, J. D. 2000, \apj, 544, L115
\bibitem[Sari, Narayan, \& Piran(1996)]{sar96} Sari, R., Narayan,
     R., \& Piran, T. 1996, \apj, 473, 204
\bibitem[Schaefer(2004)]{sch04} Schaefer, B. 2004, \apj, 602, 306
\bibitem[Shen, Song \& Li(2005)]{she05} Shen, R. F., Song, L. M.,
     \& Li, Z. 2005, \mnras, 362, 59
\bibitem[Soderberg et al.(2006)]{sod06} Soderberg, A. M., et al. 2006, Nature,
     442, 1014
\bibitem[Toma et al.(2007)]{tom07} Toma, K., Ioka, K., Sakamoto, T., \& Nakamura,
     T. 2007, \apj, 659, 1420
\bibitem[Wang et al.(2000)]{wan00} Wang, J. C., Cen, X. F., Qian, T.
     L., Xu, J., \& Wang, C. Y. 2000, \apj, 532, 267
\bibitem[Wang et al.(2006)]{wan06} Wang, X.-Y., Li, Z., Waxman, E.,
     \& Me¡äsza¡äros, P. 2006, \apj, in press (astroph/0608033)
\bibitem[Woosley \& MacFadyen(1999)]{woo99}Woosley, S. E., \& MacFadyen,
     A. I. 1999, A\&AS, 138, 499
\bibitem[Wu \& Fenimore(2000)]{wu00} Wu, B., \& Fenimore E. 2000,
     \apj, 535, L29
\bibitem[Yi et al.(2006)]{yi06} Yi, T. F., Liang, E. W., Qin, Y.
     P., \& Lu, R. J. 2006, \mnras, 367, 1751
\bibitem[Zhang \& M\'{e}sz\'{a}ros(2004)]{zha04} Zhang, B., \& M\'{e}sz\'{a}ros,
    P. 2004, Int. J. Mod. Phys. A, 19, 2385
\bibitem[Zhang et al.(2006a)]{zha06a} Zhang, B., Fan, Y. Z.; Dyks, J., Kobayashi, S.,
   M\'{e}sz\'{a}ros, P., Burrows, D. N., Nousek, J. A., \& Gehrels, N.
    2006a, \apj, 642, 354
\bibitem[Zhang \& Qin (2005)]{zha05}Zhang, F.-W., \& Qin, Y.-P. 2005, Chin. Phys., 14, 2276
\bibitem[Zhang et al.(2002)]{zha02} Zhang, Y. H., et al. 2002, \apj,
     572, 762
\bibitem[Zhang et al.(2006b)]{zha06b}Zhang, Z., Xie, G. Z., Deng, J. G., \& Jin, W. 2006b, MNRAS, 373,
729
\bibitem[Zhang et al.(2006c)]{zha06c}Zhang, Z.-B., Deng, J.-G., Lu, R.-J., \& Gao, H.-F. 2006c, ChJAA, 6, 312
\end{thebibliography}
\end{document}